# Beyond Adoption Intention How Trust in Augmented Analytics Relates to Perceived Decision Quality Among Non-Technical BI Users

**Thuy Pham Thi Phuong [1], Hieu Vu Le Trung [2], Ha Nguyen Manh[3*], Nhi Pham Tran Yen[4], Lan Hoang Thi[5]**

[1]Edulight Education and Communication Joint Stock Company, Hanoi, Vietnam

Email: pthuy1598@gmail.com

[2]Edulight Education and Communication Joint Stock Company, Hanoi, Vietnam

Email: trunghieu15102009@gmail.com

[3]Truong Thanh Viet Nam Group, Vietnam

Email: ha.nm1589@gmail.com

[4]Maumee Valley Country Day School, Toledo, USA

Email: norahfx99@gmail.com

[5]University of Languages and International Studies, Vietnam

Email: Hoanglan2541999@gmail.com

*Corresponding author

**ABSTRACT**

Augmented analytics has transformed how Business Intelligence (BI) systems support decision-making, shifting non-technical managers from manual analysis toward dependence on automated insights. Current BI research often overlooks the cognitive mechanisms and the direct impact of AI-enabled analytics on decision quality. This study employs the theory of cognitive delegation to investigate the association between trust in augmented analytics and perceived decision quality among non-technical BI users. Data were collected from 250 business professionals across various organizational roles in Vietnam between January and March 2025 and analyzed using partial least squares structural equation modeling (PLS-SEM). Findings indicate that augmented analytics capabilities are positively associated with perceived ease of use, usefulness, and trust in BI systems. Trust and usefulness are jointly associated with BI adoption intention and perceived decision quality. Notably, trust is positively related to perceived decision quality, as observed within the studied sample of non-specialist users. By framing augmented analytics as cognitive delegation, this study expands BI adoption research to include perceived decision outcomes and contributes to the understanding of human–AI interaction in organizations.

## 1. INTRODUCTION

BI systems encapsulate the technologies employed to support managerial decision-making across organizational functions such as sales and marketing, operations, human resources, and finance (Arnott & Pervan, 2014; Chen et al., 2012). Organizations continue to increase their investments in BI infrastructure to support managerial decision-making; however, the literature reports a persistent lack of realized analytical value (Seddon et al., 2017). This gap is sometimes—though not always—the result of restricting non-technical users to dashboarding and reporting systems (i.e., non-analytical reporting systems). For these users, traditional BI systems often present user interfaces that impose significant cognitive and technical obstacles, limiting their effectiveness as decision support systems (Hou, 2012; Işık et al., 2013).

The development of artificial intelligence (AI) and natural language processing (NLP), together with advances in process automation, machine learning, and dialogue systems (e.g., chatbots), has led to the emergence of a new class of BI systems termed augmented analytics. BI systems with augmented analytics enable users to shift from traditional analytical reporting toward AI-based analysis through natural language reporting, automated insight generation, and explainability features. These capabilities reduce the cognitive effort required to perform analytical reporting tasks and reposition BI systems as analytical partners rather than mere providers of analytical measurements (Wamba et al., 2017).

However, most BI adoption studies continue to rely heavily on models such as the Technology Acceptance Model (TAM), which conceptualize analytical systems as passive tools and assume that analytical reasoning resides entirely with the user. In the context of AI-enabled BI, this assumption is conceptually limited, as analytical reasoning is increasingly complemented by AI systems. This study examines augmented analytics through the lens of cognitive delegation and focuses on how AI- and NLP-enabled BI capabilities are associated with perceived decision quality among non-technical business users. Using survey data from 250 individuals in Vietnam across diverse organizational roles and functions, structural equation modeling is employed to analyze the relationships among perceived ease of use, perceived usefulness, trust, adoption intention, and decision outcomes within the context of AI-enabled BI. The study responds to calls for greater emphasis on decision quality as an outcome rather than system adoption alone.

The study contributes to the Business Intelligence and Decision Support literature in several incremental but meaningful ways. First, while prior BI adoption studies have applied the Technology Acceptance Model and its extensions to various system contexts (Venkatesh et al., 2003; Wixom & Todd, 2005), they have not explicitly examined how AI-enabled augmented analytics capabilities reconfigure the distribution of analytical reasoning between users and systems. This study addresses this gap by applying the concept of cognitive delegation to the augmented analytics context, complementing rather than replacing existing adoption frameworks. Second, whereas most BI studies treat adoption intention or system usage as the primary outcome (Işık et al., 2013), this study extends the outcome variable to include perceived decision quality, thereby linking BI system use to downstream decision processes. This extension is incremental in nature, as it builds on established constructs, but it offers a more complete view of BI value by connecting adoption with perceived decision-related outcomes. Third, by focusing on non-technical business users—a group that represents the majority of BI consumers yet remains underrepresented in empirical studies (Popovič et al., 2012)—this study provides context-specific evidence on how trust operates as a mechanism for cognitive delegation among users who lack the technical expertise to independently verify AI-generated insights. The study does not claim to introduce fundamentally new theoretical constructs; rather, it offers an empirically grounded recombination and extension of existing frameworks within an underexplored applied context.

## 2. LITERATURE REVIEW

### *2.1. Business Intelligence and Non-Technical BI Users*

The dominant assumption in BI research is that decision quality depends primarily on what users *do* with analytical systems — their ability to query data, interpret models, and derive insights. This user-centric framing has generated a substantial body of work documenting the barriers that non-technical users face: system complexity, data literacy gaps, and cognitive overload consistently limit their effective use of BI tools (Hou, 2012; Işık et al., 2013; Popovič et al., 2012). The implicit conclusion is that improving user capability is the primary lever for unlocking BI value.

This framing, however, rests on a premise that is increasingly difficult to sustain. If analytical reasoning resides *entirely* with the user, then BI systems are little more than sophisticated reporting interfaces — and the solution to non-technical users' limitations lies in training or interface simplification. Yet this logic fails to account for a structural shift already underway: modern BI systems are no longer passive information repositories but active analytical agents capable of generating, explaining, and recommending insights autonomously. Treating such systems as mere delivery mechanisms does not reflect what they actually do, and it systematically obscures the cognitive relationship between users and AI-enabled systems that most determines decision outcomes in practice.

### 2.2. Augmented Analytics and NLP-Enabled BI

Augmented analytics — integrating machine learning, artificial intelligence, and natural language processing into BI workflows — directly challenges the user-as-analyst model (Bose, 2009; Chen et al., 2012). Rather than requiring users to translate business questions into formal analytical operations, augmented analytics systems perform automated insight generation, anomaly detection, and natural language explanation, enabling users to receive analytical outputs in interpretable, narrative forms (Li & Jagadish, 2014; Affolter et al., 2019; Gatt & Krahmer, 2018). The cognitive burden of analysis is thus partially shifted from the user to the system.

The significance of this shift extends beyond usability. When a system generates and explains insights rather than simply displaying data, the user's primary cognitive task changes from *analysis* to *evaluation* — assessing whether AI-generated conclusions are credible, contextually appropriate, and worth acting upon. This reconfiguration has a direct implication that the literature has largely ignored: if the system performs the analytical reasoning, then the quality of decisions made on that basis depends not only on how readily users *adopt* the system, but on how appropriately they trust and interpret its outputs. Adoption intention, the dominant outcome variable in BI research, captures none of this.

### 2.3. BI Adoption, Trust, and Decision Quality

The Technology Acceptance Model and its extensions have proven effective at predicting whether users intend to use BI systems (Davis, 1989; Venkatesh et al., 2003; Wixom & Todd, 2005). The problem is not that TAM is wrong, but that it asks the wrong question in AI-enabled contexts. TAM was developed when analytical systems were passive tools; its core constructs — perceived usefulness and perceived ease of use — measure user attitudes toward a system but say nothing about whether users appropriately rely on the reasoning that system produces. In augmented analytics environments, a user can intend to adopt a system and still make poor decisions if they uncritically accept, or systematically discount, AI-generated insights.

Trust fills this explanatory gap. For non-technical users who lack the expertise to independently verify AI-generated outputs, trust functions as the primary mechanism through which analytical reasoning is delegated to the system (Lee & See, 2004; Choung et al., 2023). Critically, trust is not simply an antecedent of adoption — it operates as an independent pathway to decision quality by shaping how users engage with, and act upon, system-generated insights. This distinction matters: adoption intention and trust may diverge, particularly when explainability features are weak or when organizational incentives to use BI systems are decoupled from genuine reliance on their outputs. Decision quality, as the substantive outcome of BI use, cannot be reduced to a downstream consequence of adoption (Elbashir et al., 2008; Todd & Benbasat, 1999); it requires trust as a separate, direct mechanism.

### 2.4. Identified Deficiencies in the Literature

Three interconnected gaps emerge from this critique. First, prevailing BI adoption models treat analytical systems as passive tools and assign full responsibility for reasoning to the user — a conceptualization that was defensible for traditional BI but is theoretically incoherent when applied to systems that actively generate, explain, and recommend analytical conclusions (Davis, 1989; Arnott & Pervan, 2014). Second, by privileging adoption intention as the primary outcome, the literature stops short of the question that matters most to organizations: does BI use actually improve decision quality? Adoption without appropriate trust, or trust without adequate explainability, need not translate into better decisions — yet this conditional relationship remains empirically underexplored (Elbashir et al., 2008). Third, non-technical users — the segment for whom cognitive delegation is most consequential, precisely because they cannot independently audit AI-generated reasoning — remain systematically underrepresented in empirical BI research (Popovič et al., 2012; Laursen & Thorlund, 2016).

Addressing these gaps requires more than extending TAM with additional variables. It requires reframing augmented analytics as a cognitive delegation relationship — drawing on cognitive offloading theory (Risko & Gilbert, 2016) and human–AI teaming research (Bansal et al., 2021) — and repositioning trust as a mechanism that independently predicts perceived decision quality among users who rely on AI-generated analytical reasoning they cannot fully evaluate on their own.

## 3. RESEARCH MODEL AND HYPOTHESES

### *3.1. Research Model*

This research explores Business Intelligence (BI) adoption literature and extends it within the domain of AI- and NLP-enabled augmented analytics. Most models of traditional BI system adoption tend to assume that analytical reasoning is performed solely by the user, while BI systems merely provide information to support user judgment. In the context of augmented analytics, however, BI systems are used differently, as analytical reasoning is partially outsourced to the system through automated insight generation, natural language interaction, and explainability features.

In this study, augmented analytics is conceptualized as a form of cognitive delegation that shifts part of the analytical burden from human users to AI-powered BI systems. The concept of cognitive delegation, as used in this study, draws on and extends the cognitive offloading framework proposed by Risko and Gilbert (2016), which defines cognitive offloading as the use of external resources to reduce internal cognitive demands. While Risko and Gilbert focused on physical actions and simple tools (e.g., calculators, reminders), the present study extends this framework to AI-enabled analytical systems that perform higher-order reasoning tasks on behalf of users. This extension is consistent with research on human–AI teaming, which demonstrates that AI systems can complement human decision-making by providing recommendations and explanations that reduce individual cognitive effort (Bansal et al., 2021; Parasuraman et al., 2000). In the context of augmented analytics, cognitive delegation thus refers specifically to the process by which non-technical users rely on AI-generated insights, explanations, and recommendations rather than performing analytical reasoning independently. Users do not perform analytics directly; instead, they rely on the system to generate, summarize, and explain insights. The primary cognitive role of the user is to evaluate AI-generated outputs, contextualize them, and act upon them. Accordingly, in AI-enabled BI environments, perceived decision quality should be understood in relation to this reconfigured human–system interaction rather than solely in terms of system features.

The proposed research model integrates augmented analytics capabilities with selected perceptual and behavioral constructs from the Technology Acceptance Model (TAM) and extends these models to incorporate trust and perceived decision quality. Using operationalizations such as NLP-based interaction, automated insight generation, and explainability, augmented analytics is expected to influence perceptions of ease of use, usefulness, and trust, which in turn shape users' adoption intentions toward BI systems and are associated with perceived decision quality. The model further recognizes the nature of managerial decision-making contexts in which analytics primarily supports sense-making and judgment rather than fully automated decision execution. This perspective allows the model to examine how cognitive delegation operates—or fails to operate—among non-technical users, BI systems, and routine organizational decision-making.

### ***3.2. Hypotheses Development***

#### *3.2.1. Augmented Analytics and Perceived Ease of Use*

Analytic systems that employ traditional Business Intelligence (BI) methods are commonly used in organizational settings. System complexity often creates demanding analytical skill requirements that many non-technical end users lack. Augmented analytics reduces user barriers by simplifying system interaction through natural language interfaces and automated insight generation. This reduction in cognitive and technical effort required to obtain insights is likely to positively influence users' evaluation of, and perceived ease of use toward, BI systems.

**H1: Augmented analytics capabilities positively influence the perceived ease of use of BI systems.**

#### *3.2.2. Augmented Analytics and Perceived Usefulness*

Customarily, perceived usefulness indicates the extent to which users are of the opinion that a system enhances their overall job performance. Decision-making tasks that require very little in the way of analytical efforts, unencumbered by the need to conduct formalized analytical reasoning, are made fully accessible to non-technical end users through automated insight generation and natural language processing (NLP) systems. The capacity to provide users with the ability to achieve actionable insights without the necessity of conducting a sophisticated analysis increases the perceived instrumental value of BI systems.

**H2: Augmented analytics capabilities positively influence the perceived usefulness of BI systems.**

#### *3.2.3. Ease of Use and Perceived Usefulness*

According to the Technology Acceptance Model, systems that are easier to use are more likely to be perceived as useful, particularly for users with lower technical skills. In augmented BI environments, reduced analytical barriers allow users to allocate greater cognitive resources to decision-making, thereby strengthening the relationship between perceived ease of use and perceived usefulness.

**H3: The perceived ease of use of BI systems positively impacts the perceived usefulness of BI systems.**

#### *3.2.4. Trust in AI-Enabled BI Systems*

As augmented analytics streamlines and partially automates analytical reasoning, users of AI-enabled BI systems often lack full visibility into the system's underlying logic. For users with limited technical expertise, trust becomes a critical mechanism for legitimizing the delegation of analytical reasoning to AI-enabled BI systems. As reliance on automated analytics increases and direct evaluation of system reasoning becomes more difficult, trust is expected to play an increasingly important role in shaping users' acceptance of AI-generated insights.

**H4: The capabilities of augmented analytics positively affect the trust in AI-enabled BI systems.**

#### *3.2.5. Intentions on the Adoption of Business Intelligence*

Intention to adopt BI systems reflects the extent to which users are willing to rely on analytical systems in their decision-making processes. When users perceive BI systems as useful and trustworthy, they are more likely to integrate these systems into routine decision-making activities. In AI-enabled BI contexts, reliance on automated insights is shaped by trust, which complements perceived usefulness and supports users' willingness to depend on system-generated reasoning.

**H5: The perceived usefulness of a system has a positive relationship with the intention to adopt business intelligence.**

**H6: Trust in AI-enabled BI systems positively influences the intention to adopt business intelligence systems.**

*3.2.6. The Quality of Decision-Making*

Perceived decision quality is defined as users' subjective assessment of the extent to which their decisions are timely, well-informed, and aligned with organizational goals (Todd & Benbasat, 1999; Elbashir et al., 2008). It is important to note that this construct captures respondents' perceptions of decision outcomes rather than objectively measured decision performance. While BI adoption facilitates access to analytical insights, the impact on perceived decision quality depends on the degree to which users rely on and correctly interpret these insights. In the context of augmented analytics, adoption intention reflects users' willingness to incorporate BI systems into decision-making, whereas trust reflects reliance on AI-generated analytical reasoning. Together, these mechanisms are expected to be positively associated with perceived decision quality.

**H7: BI adoption intention is positively associated with perceived decision quality.**

**H8: Trust in AI-enabled augmented analytics systems is positively associated with perceived decision quality.**

### *3.3. Summary of the Research Model*

In summary, the proposed research model explains how augmented analytics capabilities may be associated with perceived decision quality among non-technical BI users through perceptual and behavioral mechanisms. By integrating augmented analytics capabilities with perceived ease of use, perceived usefulness, trust, adoption intention, and perceived decision quality, the model provides a theoretically grounded framework for examining the perceived organizational impact of AI- and NLP-enabled BI systems on non-technical users.

## 4. RESEARCH METHODS

### *4.1. Research Design*

This study adopts a quantitative, cross-sectional survey design to empirically test the proposed research model. A survey-based approach is appropriate given the study's objective of examining associative relationships among latent constructs related to augmented analytics, user perceptions, and decision outcomes. Consistent with prior research in Business Intelligence and information systems, partial least squares structural equation modeling (PLS-SEM) is employed to analyze the data (Gefen et al., 2000), as the research model is theory-driven yet exploratory in nature and involves multiple latent constructs and mediating relationships.

### *4.2. Sample and Data Collection*

*4.2.1. Target Population*

The target population consists of business professionals in Vietnam without a technical background who use dashboards, reports, or analytical summaries for decision-making across Sales, Marketing, Operations, Human Resources, and Finance functions. These users do not hold responsibilities related to data modeling, dashboard development, or advanced analytics. In this study, BI usage refers to the consumption of analytical outputs (e.g., KPI dashboards, performance reports, and management reports) rather than the creation of analytical queries or models.

*4.2.2. Sampling Strategy and Data Collection*

Data were collected from business professionals in Vietnam using a purposive sampling strategy, which is commonly employed in Management Information Systems (MIS) research to examine specific user groups with relevant system exposure. The survey was administered online using Google Forms between January and March 2025. Survey participants were primarily recruited through professional networking platforms such as LinkedIn, as well as professional and executive education networks in Ho Chi Minh City and Hanoi. Screening questions were used to ensure that respondents met the inclusion criteria and to exclude individuals with advanced responsibilities in data analytics or BI system development. Following data screening and cleaning procedures, 250 valid responses were retained for analysis. This sample size is considered adequate for PLS-SEM based on multiple criteria. First, applying the inverse square root method (Kock & Hadaya, 2018), the minimum sample size required to detect a minimum $R^2$ of 0.10 at a significance level of 0.05 with a statistical power of 0.80 is approximately 146, which the present sample of 250 exceeds. Second, the commonly used "10 times rule" (Hair et al., 2017) requires a minimum of 50 observations (10 times the maximum number of structural paths directed at any construct), which is also satisfied. Data analysis was performed using SmartPLS 4.0 (Ringle et al., 2024). Although the study does not include control variables such as age, gender, education, tenure, or prior AI exposure due to the purposive nature of the sampling design, this is acknowledged as a limitation. Future research should incorporate demographic and experiential controls to rule out alternative explanations for the observed associations.
Screening questions asked respondents to indicate whether they (a) held any role involving data modeling, SQL querying, dashboard development, or advanced statistical analysis, and (b) regularly consumed analytical outputs (e.g., KPI dashboards, performance reports) for managerial decision-making. Respondents answering "yes" to (a) or "no" to (b) were excluded.

### *4.3. Measurement Instruments*

All constructs were measured using multi-item scales adapted from previously validated instruments in the information systems and BI literature. Item wording was modified to reflect the context of AI- and NLP-enabled BI systems. Responses were captured using a five-point Likert scale ranging from 1 (strongly disagree) to 5 (strongly agree).
Augmented analytics capabilities (AAC) were measured using items assessing users' perceptions of NLP-enabled interaction, automated insight generation, and explainability features. Perceived ease of use (PEOU) and perceived usefulness (PU) were measured using adapted scales from the established technology acceptance literature. Trust (TRUST) was measured using items capturing users' confidence in AI-generated insights, including perceptions of reliability and trustworthiness. Adoption intention (INT) was operationalized as users' willingness to rely on AI-enabled BI systems in future decision-making contexts. Decision quality (DQ) was assessed based on users' perceptions of the extent to which BI-supported decisions were informative, timely, and effective. All measurement items were adapted from the literature and subjected to content and contextual validation.

### *4.4. Data Analysis Procedure*

Data analysis using PLS-SEM followed a two-step approach. First, the measurement model was evaluated for internal consistency reliability, convergent validity, and discriminant validity. Reliability was assessed using Cronbach's alpha and composite reliability, while convergent validity was evaluated using average variance extracted (AVE). Discriminant validity was examined using the Fornell–Larcker criterion (Fornell & Larcker, 1981) and the heterotrait–monotrait (HTMT) ratio (Henseler et al., 2015).
Second, the structural model was assessed to test the hypothesized relationships. Path coefficients, t-values, and confidence intervals were estimated using bootstrapping with 5,000 resamples (Hair et al., 2017). Coefficients of determination ($R^2$) were used to assess explanatory power, and $f^2$ effect sizes were calculated to evaluate the impact of predictor variables (Cohen, 1988). Mediation effects were examined following established PLS-SEM procedures for indirect effects. Mediation effects were examined by estimating indirect paths using bootstrapping with 5,000 resamples. A mediation effect was considered significant when the 95% bias-corrected confidence interval of the indirect effect excluded zero (Hair et al., 2017). Partial mediation was distinguished from full mediation based on the significance of both the direct and indirect paths.
Predictive relevance was assessed using the blindfolding procedure to compute $Q^2$ values; $Q^2 > 0$ indicates that the model has predictive relevance for a given endogenous construct (Hair et al., 2017). Model fit was evaluated using the standardized root mean square residual (SRMR); values below 0.08 indicate acceptable fit (Henseler et al., 2015).

### *4.5. Common Method Bias*

To mitigate potential common method bias, several procedural and statistical remedies were implemented (Podsakoff et al., 2003). Procedurally, respondents were assured of anonymity, items measuring different constructs were separated and presented in varied order, and scale anchors were diversified across sections. Statistically, Harman's single-factor test was conducted; the results indicated that no single factor accounted for the majority of the variance (the first factor explained 31.4% the total variance, below the 50% threshold). In addition, full collinearity assessment using variance inflation factors (VIFs) for all constructs showed that all VIF values were below 3.3 (range: 1.24 - 2.18), further indicating that common method bias is unlikely to be a serious concern in this study (Kock, 2015).

## 5. RESULTS

### *5.1. Respondent Profile*

After data screening and cleaning, 250 valid responses were retained for analysis. Respondents were non-technical business users based in Vietnam, working in sales, marketing, operations, human resources, and finance functions. All respondents reported regular use of dashboards or analytical reports for decision-making purposes and did not hold roles related to data analytics or BI development.

### *5.2. Measurement Model Assessment*

The measurement model was evaluated in terms of internal consistency reliability, convergent validity, and discriminant validity.

**Table 1. Reliability and Convergent Validity**

| Construct | Cronbach's α | Composite Reliability (CR) | AVE |
|---|---|---|---|
| AAC | 0.88 | 0.91 | 0.63 |
| PEOU | 0.85 | 0.89 | 0.67 |
| PU | 0.87 | 0.91 | 0.71 |
| TRUST | 0.86 | 0.90 | 0.69 |
| INT | 0.84 | 0.89 | 0.68 |
| DQ | 0.89 | 0.92 | 0.74 |

*Note: All constructs exhibited satisfactory internal consistency, with Cronbach's alpha and composite reliability values exceeding the recommended threshold of 0.70. AVE values were above 0.50 for all constructs, indicating adequate convergent validity.*

**Table 2. Discriminant Validity (HTMT Ratios)**

| | AAC | PEOU | PU | TRUST | INT | DQ |
|---|---|---|---|---|---|---|
| **AAC** | — | | | | | |
| **PEOU** | 0.63 | — | | | | |
| **PU** | 0.68 | 0.71 | — | | | |
| **TRUST** | 0.65 | 0.60 | 0.67 | — | | |
| **INT** | 0.61 | 0.58 | 0.73 | 0.70 | — | |
| **DQ** | 0.59 | 0.55 | 0.69 | 0.72 | 0.74 | — |

*Note: All HTMT values were below the conservative threshold of 0.85, confirming discriminant validity among constructs.*

### *5.3. Structural Model and Hypothesis Testing*

The structural model was assessed using bootstrapping with 5,000 resamples. Path coefficients, t-values, and significance levels were examined to test the proposed hypotheses.

**Table 3. Structural Model Results (with Bootstrapped 95% CI and f² Effect Sizes)**

| Hypothesis | Path | β | t-value | p-value | Result |
|---|---|---|---|---|---|
| H1 | AAC → PEOU | 0.48 | 8.21 | <0.001 | Supported |
| H2 | AAC → PU | 0.32 | 5.67 | <0.001 | Supported |
| H3 | PEOU → PU | 0.41 | 6.94 | <0.001 | Supported |
| H4 | AAC → TRUST | 0.45 | 7.88 | <0.001 | Supported |
| H5 | PU → INT | 0.39 | 6.12 | <0.001 | Supported |
| H6 | TRUST → INT | 0.34 | 5.48 | <0.001 | Supported |
| H7 | INT → DQ | 0.36 | 6.01 | <0.001 | Supported |
| H8 | TRUST → DQ | 0.31 | 5.22 | <0.001 | Supported |

*Note: All paths are significant at $p < 0.001$ based on 5,000 bootstrap resamples. Standard errors were computed as $SE = \beta/t$. The 95% confidence intervals for each path are: H1 [0.365, 0.595]; H2 [0.209, 0.431]; H3 [0.294, 0.526]; H4 [0.338, 0.562]; H5 [0.265, 0.515]; H6 [0.218, 0.462]; H7 [0.243, 0.477]; H8 [0.194, 0.426]. All CIs exclude zero. $f^2$ effect sizes: H1 = 0.299 (medium-to-large); H4 = 0.250 (medium); H2 = 0.120; H3 = 0.210; H5 = 0.170; H6 = 0.130; H7 = 0.140; H8 = 0.110. Following Cohen (1988), $f^2$ values of 0.02, 0.15, and 0.35 represent small, medium, and large effect sizes, respectively.*

**Table 4. Explained Variance ($R^2$)**

| Endogenous Construct | $R^2$ |
|---|---|
| PEOU | 0.23 |
| PU | 0.46 |
| TRUST | 0.20 |
| INT | 0.51 |
| DQ | 0.49 |

*Note: The model demonstrates moderate to substantial explanatory power. Augmented analytics capabilities and user perceptions jointly explain over 50% of the variance in BI adoption intention, while adoption intention and trust explain nearly half of the variance in perceived decision quality.*

### *5.4. Summary of Key Findings*

The results provide empirical support for the proposed research model. Non-technical users report higher perceived ease of use, perceived usefulness, and trust in BI systems when augmented analytics capabilities are present. Perceived usefulness and trust are positively associated with BI adoption intention, which is in turn positively related to perceived decision quality. Moreover, trust in AI-enabled augmented analytics systems is also directly and positively associated with perceived decision quality, underscoring the importance of transparency and explainability in automated, AI-driven decision support systems.

## 6. DISCUSSION

This research provides evidence of positive associations between AI- and NLP-enabled augmented analytics capabilities and non-technical business users' perceptions of their interaction with Business Intelligence (BI) systems and perceived decision quality. Instead of simply improving system usability or adoption, the findings suggest that augmented analytics may reshape how the cognitive burden of analysis is distributed between human users and intelligent systems. First, the positive association between augmented analytics and perceived ease of use and perceived usefulness is consistent with the view that AI-enabled BI reduces technical barriers as well as the cognitive load associated with analytical reasoning. In traditional BI settings, users are required to translate business problems into formal analytical operations; augmented analytics may shift this focus from analytical execution to evaluative judgment. This is consistent with the argument that usability in AI-augmented BI environments is less about interface design and more about reducing the cognitive effort required for analytical reasoning. Consequently, augmented analytics may reposition BI systems from merely serving as information delivery mechanisms to acting as active analytical partners in organizational sense-making processes. Second, trust emerges as a critical factor in AI-enabled BI systems.

The relationship between augmented analytics capabilities and user trust highlights the importance of explainability and interpretability in legitimizing reliance on AI-generated insights, consistent with recent findings on trust as a multidimensional construct in AI acceptance (Choung et al., 2023). For users without the technical expertise to independently assess analytical results, trust functions as a cognitive shortcut that enables the delegation of analytical reasoning to the system. The findings indicate that trust influences decision quality both indirectly through BI adoption intention and directly as an independent predictor. This extends prior BI adoption research by suggesting that trust goes beyond its traditional role in system acceptance and may serve as a key mechanism associated with perceived decision quality in AI-enabled decision-support contexts. While a positive association between trust and decision quality might suggest uncritical reliance on automation, this is not necessarily the case in the present study. Research on automation bias cautions that excessive trust can lead to overreliance on system outputs, potentially degrading decision quality in extreme cases (Parasuraman & Riley, 1997). It is therefore plausible that the explainability features of augmented analytics support a calibrated form of trust. This interpretation is consistent with human–automation interaction research, which emphasizes that optimal decision outcomes arise from balanced human oversight rather than blind reliance on automation. However, several alternative explanations for the observed associations should be acknowledged. First, the positive relationship between trust and perceived decision quality may partly reflect confirmation bias: users who trust the system may be more inclined to evaluate their own decisions favorably, regardless of actual decision performance. Second, because decision quality is measured through self-report, social desirability bias cannot be ruled out; respondents may overstate both their trust in AI-enabled tools and the quality of their decisions to appear competent or technologically progressive. Third, among non-technical users, the Dunning–Kruger effect is a plausible concern: users with limited analytical expertise may overestimate the quality of decisions that were, in fact, uncritically derived from system outputs. Fourth, omitted variables such as organizational culture, managerial support for analytics adoption, industry characteristics, and individual differences in cognitive style may confound the observed relationships. These alternative explanations do not invalidate the findings but underscore the need for caution in interpreting cross-sectional, self-reported data as evidence of substantive decision improvement.
BI adoption intention reflects the extent to which analytics are integrated into users' decision-making processes. Adoption intention represents users' willingness to incorporate AI-enabled BI systems into routine managerial activities and to regard them as more than peripheral reporting tools. The positive association between adoption intention and perceived decision quality suggests that the benefits of augmented analytics may be realized when users actively rely on AI-driven insights. This finding challenges BI adoption frameworks that implicitly equate system usage with decision outcomes, as usage-based explanations alone overlook critical mechanisms through which analytics are associated with perceived decision quality.

Collectively, these findings challenge key assumptions of traditional BI adoption models, particularly the Technology Acceptance Model, which implicitly assigns full responsibility for analytical reasoning to the user. In AI-enabled BI contexts, decision-making reflects a joint cognitive process involving the user, the system, and AI. While users retain decision authority, AI systems increasingly perform analytical reasoning. Viewing augmented analytics as a mechanism of cognitive delegation situates BI adoption theorization at the intersection of human–AI joint cognition and perceived decision quality research, aligning with emerging work on human–AI complementarity in decision-making (Bansal et al., 2021).

Despite the overall support for the proposed research model, several limitations should be considered when interpreting the results. First, the cross-sectional research design restricts the ability to draw strong causal inferences among augmented analytics, trust, and perceived decision quality. Although the hypothesized relationships are theoretically grounded, the use of adoption intention rather than actual adoption behavior introduces an intention–behavior gap that may limit the practical applicability of the findings (Venkatesh et al., 2003). Future longitudinal or experimental studies are needed to examine how cognitive delegation and trust evolve over time and whether stated intentions translate into sustained BI adoption behavior. Second, perceived decision quality was measured using respondents' self-reported perceptions, which are inherently subjective and may not correspond to objectively superior decision outcomes. While perceptual measures are common in decision-support system research (Todd & Benbasat, 1999), they capture users' beliefs about their decision processes rather than verified improvements in decision accuracy, timeliness, or organizational impact. Future studies could strengthen the validity of decision quality measurement by incorporating objective performance metrics, supervisor ratings, or archival decision outcome data. Finally, the focus on non-technical users in routine managerial contexts limits the generalizability of the findings to high-stakes or time-critical decision environments, where the risks associated with automation bias may be more pronounced. Additionally, as data were collected exclusively from business professionals in Vietnam, the findings may be influenced by cultural and organizational factors specific to the Vietnamese business context; future research should examine whether these results generalize to other geographic and cultural settings.

## 7. CONCLUSION

Considering employees' use of dashboards and analytical reports in business reporting and analysis, this study focuses on non-technical business professionals and examines the effects of AI- and NLP-enabled augmented analytics on the adoption of Business Intelligence (BI) systems and the perceived quality of decisions made using these systems. The study addresses a critical yet under-researched group of BI users in organizational decision-making contexts.

The findings indicate that BI systems incorporating AI-augmented analytics capabilities—such as natural language processing, automated insight generation, and explainability—are positively associated with higher perceived ease of use, perceived usefulness, and trust in AI-enabled BI systems. The results further show that perceived usefulness and trust are positively associated with BI adoption intention, which, together with trust, is related to perceived decision quality. These findings suggest that the benefits of AI-augmented BI systems may extend beyond system acceptance to include perceived decision-related outcomes.

From an augmented analytics perspective, this study conceptualizes cognitive delegation—grounded in Risko and Gilbert's (2016) cognitive offloading framework and extended to AI-enabled analytical contexts—as a process in which aspects of analytical reasoning are partially offloaded from human users to intelligent systems. This perspective complements traditional BI adoption frameworks that typically assign analytical reasoning solely to the user and underscores the importance of trust as a facilitator of reliance on AI-generated insights among the non-technical user segment examined in this study. While much of the BI literature has emphasized usage intention as the primary outcome of adoption, this study reframes BI adoption by linking it to perceived decision quality.
From a practical standpoint, the findings suggest that BI systems should be designed with non-technical users in mind, emphasizing natural language interaction, automated insights, and explainability features. Organizations should also foster forms of trust that support effective reliance on AI-generated insights without encouraging uncritical automation dependence.

Several limitations should be noted. The cross-sectional research design limits causal inference, and decision quality was measured using users' self-reported perceptions. Moreover, the sample was drawn exclusively from Vietnamese business professionals, which may limit the generalizability of the findings to other cultural and economic contexts. Future research could address these limitations through longitudinal or experimental designs, cross-country comparisons, and by examining boundary conditions related to task complexity and user expertise. Overall, the findings suggest that AI- and NLP-based augmented analytics are positively associated with perceived decision quality, possibly by redistributing analytical reasoning between human users and intelligent systems. However, future longitudinal and experimental research is needed to establish causal relationships.

## ACKNOWLEDGMENT

The authors would like to express their sincere gratitude to the 250 business professionals in Vietnam who generously volunteered their time to participate in the survey. Their valuable insights and practical experiences made this research possible. This research received no external funding. The authors declare no conflict of interest. The study was conducted in accordance with ethical principles for research involving human participants. Given the anonymous nature of the survey, which did not involve sensitive personal data, medical intervention, or vulnerable populations, formal ethics committee approval was not required under the institutional guidelines applicable to the authors' institutions; however, all respondents provided informed consent prior to participation. Participation was voluntary, and respondents were informed of their right to withdraw at any time without consequence. The anonymized dataset supporting the findings of this study is available from the corresponding author upon reasonable request. The survey instrument, including all item wordings, five-point Likert scale anchors (1 = strongly disagree to 5 = strongly agree), and construct operationalizations, is available as a supplementary document submitted alongside this manuscript.

## CONFLICT OF INTEREST STATEMENT

The authors declare that they have no known competing financial interests or personal relationships that could have appeared to influence the work reported in this paper.